\documentclass[showpacs,twocolumn]{revtex4-1}
\usepackage[english]{babel}
\usepackage{graphicx}
\usepackage{epstopdf}
\usepackage{amsmath, amssymb}
\usepackage{dcolumn}
\usepackage{subfigure}
\usepackage{lipsum}
\usepackage{amssymb}
\usepackage{amsmath}
\usepackage{ntheorem}

\newcommand{\Tr}{\mathrm{Tr}}

\renewcommand{\Re}{\mathrm{Re}}

\newcommand{\ket}[1]{
	\left|#1\right\rangle
}

\newcommand{\bra}[1]{
	\left\langle#1\right|
}

\newcommand{\Next}{\\[0.5cm]}
\newcommand{\cosec}{\mathrm{cosec}}
\newtheorem*{sld}{SLD theorem}
\everymath{\displaystyle}

\begin{document}

\title{\bf Quantum Estimation in Neutrino Oscillations}

\author{Edson C. Nogueira $^a$}\email[]{edsoncezar16@gmail.com}
\author{Gustavo de Souza $^b$}\email[]{gdesouza@iceb.ufop.br}
\author{Adalberto D. Varizi $^a$}\email[]{adalbertovarizi@gmail.com}
\author{Marcos D. Sampaio $^a$}\email[]{msampaio@fisica.ufmg.br}

\affiliation{$^a$ Universidade Federal de Minas Gerais -- Departamento de F\'{\i}sica -- ICEX \\ P.O. BOX 702,
	30.161-970, Belo Horizonte/MG -- Brazil,}
\affiliation{$^b$ Universidade Federal de Ouro Preto -- Departamento de Matem\'{a}tica -- ICEB \\
	Campus Morro do Cruzeiro, s/n, 35400-000, Ouro Preto/MG -- Brazil}

\begin{abstract}
Neutrino oscillations are at the forefront of advances in Physics beyond the Standard Model. Increasing accuracy in measurements of the neutrino mixing matrix is an important challenge in current experiments.~It depends on parameters that do not directly correspond to observables of the neutrino system.~This type of estimation problem is handled by Quantum Estimation Theory (QET) via the Fisher Information (FI) and the Quantum Fisher Information (QFI).~In this work, we analyze two-flavor neutrino oscillations within the framework of QET.~We compute the QFI for the mixing angle $\theta$ and show that mass measurements are the ones that achieve optimal precision.~We also study the FI associated with flavor measurements and show that they are optimized at specific neutrino times-of-flight.~Therefore, although the usual population measurement does not realize the precision limit set by the QFI, it can in principle be implemented with the best possible sensitivity to $\theta $.~We study how these quantifiers relate to the single-particle, mode entanglement. We demonstrate that this form of entanglement does not enhance neither of them.~In particular, this shows that in single-particle settings, entanglement is not directly connected with the optimal precision in metrological tasks.
\end{abstract}

\maketitle

\section{Introduction}
\label{sec:intro}

\par Neutrinos have been a unique source of fundamental discoveries in Physics since their introduction by Pauli and the development of the Fermi theory of beta decay \cite{giunti_2007, fermi_1934}. In the Standard Model of elementary particles they are massless \cite{maggiore_2005}, but the phenomena of neutrino oscillations -- proposed for the first time in 1957 by B. Pontecorvo \cite{pontecorvo_1957} and confirmed some decades later by the series of experiments that culminated in \cite{ahmad_2002} --, show that they must have mass.

\par The theory of neutrino mixing was developed in analogy with that of quark mixing. After the initial efforts of Pontecorvo \cite{pontecorvo_1957}, Maki, Nakagawa and Sakata gave relevant contributions \cite{maki_1962, pontecorvo_1969} and the theory was completed in the 70's by Bilenky and others \cite{giunti_2007, bilenky_1976}. The point is that the mismatch between flavor and mass eigenstates is described by a mixing matrix $U$ which contains some parameters -- mixing angles, neutrino masses and CP violating phases -- that encompass all the information about oscillation phenomena. Hence, a precise knowledge about the values of these parameters is essential to the comprehension of neutrino mixing and future developments in this field.

\par However, these parameters do not directly correspond to a physical observable of the neutrino system. Therefore, in order to measure them one shall infer their  values indirectly  from the measurement of other observables. For example: in neutrino oscillations experiments, population measurements are performed and the values of the mixing angles are estimated based on their outcomes. Although some parameters of the mixing matrix have been measured over the last years \cite{olive_2014, samitz_2012}, to enhance the precision on their values -- i.e., to reduce uncertainty intervals -- remains a fundamental task, and experiments are being designed to do this in the following few years \cite{dune_2015}.

\par From these considerations, it follows that it is pertinent to ask the following questions: what is the optimal precision we can achieve through population measurements in neutrino oscillations experiments? What is the ultimate precision limit allowed by the laws of Quantum Mechanics? Is it possible to determine the observable that we should measure in order to obtain it? These questions are properly addressed in the framework of Quantum Estimation Theory (QET) \cite{helstrom_1976, cramer_1999}, where we can determine the optimal precision that can possibly be achieved with any (generalized) measurement scheme through the so-called Quantum Fisher Information (QFI), and sometimes, also the measurement scheme that permits such precision to be achieved. It necessarily corresponds from a  mathematical viewpoint to the Positive Operator Valued Measure (POVM) generated by projectors over the eigenspaces of a certain operator -- the so-called Symmetric Logarithmic Derivative (SLD) \cite{toth_2014}. However, it is not always possible to interpret these projectors as physically meaningful observables.

\par In the present work, we consider two-flavor neutrino oscillations within the framework of QET and analyze the estimation of the mixing angle $\theta$ which characterizes them. We will consider neutrinos produced in a definite flavor and two types of models for the ensuing neutrino oscillations:

\vspace{-0.2cm}
\begin{enumerate}
\item[1)] the standard plane-wave model;
\item[2)] a ``decoherence model'' which takes into account the essential feature of wave-packet models, which is the dynamical suppression of coherences between different neutrino mass states and the corresponding damping of the flavor oscillation probability \cite{Bittencourt_EPL}.
\end{enumerate}

\noindent In both cases, we will calculate the Quantum Fisher Information (QFI) and the population measurement Fisher Information (PMFI) (i.e., the FI for flavor detection), which quantifies the sensibility of this commonly used protocol for the estimation of $\theta$ and the maximum precision achievable with it.~We will see that for models of type (2), which is actually the relevant one for oscillation experiments, the projectors onto eigenstates of the SLD can be physically interpreted, and direct mass measurement will be shown to be the scheme which realizes the QFI.~Moreover, we will show that the PMFI is optimized for both models at specific neutrino times-of-flight.~Therefore, although the usual flavor measurement is not the one which achieves the optimal precision set by the QFI, it can in principle be implemented with the best possible sensitivity to the desired parameter $\theta $.

\par Another point that deserves attention is entanglement. Recent works (see ref. \cite{toth_2014} for a summary) have shown, for example, that the use of entangled probe states in parameter estimation tasks make it frequently possible to reach the so-called Heisenberg-scaling, which means that the variance of the estimator of the desired parameters scales as $\mathrm{var}(\theta)\sim \tfrac{1}{N^2}$, $N$ being the number of trials. This is better than the usual shot-noise scaling  $\mathrm{var}(\theta)\sim \tfrac{1}{N}$: it gives smaller measurement uncertainty, for fewer trials. This type of result would be very interesting in contexts such as the present one of neutrino oscillations, where the trial events -- i.e., neutrino detections -- are difficult.

\par More importantly, these papers show that the use of entangled probe states implies on the enhancement of the maximum precision limit achievable, optimizing the value of the QFI. When a physical interpretation of the SLD is available, it is thus possible to use the entanglement as a resource leading to higher-performance parameter estimation. However, these results were obtained in the context of multiparticle (usually two-particle) probe systems. This makes it hard to apply them to neutrino studies: since neutrinos hardly interact one with each other in oscillations experiments, usable entangled neutrino pairs would be very hard to obtain in this context.

\par Having this in mind, we will also investigate here the role of the single-particle, mode entanglement of Blasone et al. \cite{Blasone_PRD} on the two-flavor mixing angle estimation. Due to the ``hardly interacting'' character of neutrinos mentioned above, this is the natural form of entanglement to consider for them (also referred to in the literature as occupation number entanglement \cite{Cunha_Vedral}). We investigate how it relates to the QFI and the FI for direct flavor detection. We show that this entanglement does not enhance neither of these quantifiers. In particular, this shows that at least in single-particle settings, entanglement is not directly connected with the optimal precision in metrological tasks.

\par This work is organized as follows.~In the preliminary Section 2, we review the basics of QET and set our notation for two-flavor neutrino oscillations in the plane-wave approach.~We also discuss occupation number entanglement in neutrino oscillations (following Blasone et. al \cite{Blasone_PRD}).~Next, we consider in Section 3 the QFI and SLD for model (1) above.~We will see that in this case there is no clear physical interpretation for the projectors onto eigenspaces of the SLD, in such a way that it is not possible to use them to determine an optimal measurement protocol for the mixing angle. In Section 4, we show that this is no longer the case when one takes into account the dynamical suppression of quantum coherence terms between the different mass eigenstates that compose the initial state.~This is the main Section of the present work, since this effect always occurs to some extent in oscillation experiments.~We model it by introducing in the plane-wave model a Markovian decoherence term.~It will be given by a Lindblad operator that projects into one of the mass eigenstates, therefore damping the coherences between them.~Solving the associated Lindblad master equation, we show that the QFI is the same found for model (1), but now the measurement scheme that allows the optimal precision \emph{can} be determined: it consists of the mass measurement.~The quantitative relationship between entanglement and the Fisher Information will be analyzed for each type of model in the Section dedicated to it.~We also calculate the PMFI in each case and characterize its optimization.~Finally, Section 5 summarizes the results and closes with a few concluding remarks.

\vspace{-0.2cm}
\section{Quantum estimation theory and neutrino oscillations}
\label{sec:sec2}
\subsection{Quantum estimation theory}\label{sec.qe}

\vspace{-0.2cm}
\par Estimation theory is necessary when we want to know the value of a parameter of a physical system that does not directly correspond to a physical observable \cite{paris_2009}. In this case, we make some measurements on the system and infer the value of the parameter based on the measurement outcomes. These problems are properly treated in the framework of Quantum Estimation Theory (QET) \cite{helstrom_1976}. Basically, for a given set of measurements outcomes $\chi$, we look for an \emph{estimator} $\hat{\lambda}(\chi)$ that maps every possible outcome $\chi$ to a value $\lambda=\hat{\lambda}(\chi)$ of the parameter we want to estimate. A result of estimation theory is that the variance of \emph{any} estimator $\hat{\lambda}$ is bounded from below according to the Cramer-Rao inequality \cite{cramer_1999}

\vspace{-0.3cm}
\begin{equation}
Var(\hat{\lambda})\geq \frac{1}{MF(\lambda)}, \label{eq.cramer}
\end{equation}

\noindent where $M$ denotes the number of measurements and $F(\lambda)$ is the \emph{Fisher Information} (FI):

\vspace{-0.3cm}
\begin{equation}
F(\lambda)\equiv\sum_x \frac{1}{p(x|\lambda)}\left(\frac{\partial p(x|\lambda)}{\partial\lambda}\right)^2. \label{eq.fisher}
\end{equation}

\noindent The sum runs over all possible measurement outcomes and $p(x|\lambda)$ is the conditional probability of obtaining the outcome $x$ given that the value of the parameter we want to estimate is $\lambda$.

\par In Quantum Mechanics, every information property of a given measurement can be completely described by a set $\{x, \Pi_x\}$ of measurement outcomes $x$ and non-negative operators $\Pi_x$, such that the probability to get the value $x$ is $\Tr[\Pi_x\rho]$. Here, $\rho$ denotes the density matrix of the system of interest \cite{jacobs_2014}. Such sets are called Positive Operator Valued Measures (POVM's). When the parameter has the value $\lambda$, the system is described by a density matrix $\rho_\lambda$ and we have

\vspace{-0.3cm}
\begin{equation}
p(x|\lambda)=\Tr[\rho_\lambda\Pi_x].\label{eq.pxl} 
\end{equation}

\par We define the \emph{Symmetric Logarithmic Derivative} (SLD) $L_\lambda$ by the equation

\vspace{-0.3cm}
\begin{equation}
\partial_\lambda\rho_\lambda\equiv \frac{1}{2}\left\{\rho_\lambda , L_\lambda\right\}.\label{eq.sld}
\end{equation}

\noindent By substituting eqs. \eqref{eq.pxl} and \eqref{eq.sld} into \eqref{eq.fisher}, the Fisher information reads

\vspace{-0.3cm}
\begin{equation}
F(\lambda)=\sum_x \frac{\left[\Re\;\Tr(\rho_\lambda\Pi_x L_\lambda)\right]^2}{\Tr[\rho_\lambda\Pi_x]}.\label{eq.FI}
\end{equation}

\noindent From the above equation we finally get \cite{ paris_2009, braunstein_1994}:

\vspace{-0.3cm}
\begin{align}
F(\lambda)&=\sum_x \frac{\left[\Re\;\Tr(\rho_\lambda\Pi_x L_\lambda)\right]^2}{\Tr[\rho_\lambda\Pi_x]}\nonumber\\
&\leq \left|\frac{\Tr[\rho_\lambda\Pi_x L_\lambda]}{\sqrt{\Tr[\rho_\lambda\Pi_x]}}\right|^2\label{eq.ineq1}\\
&= \left|\Tr\left[\frac{\sqrt{\rho_\lambda}\sqrt{\Pi_x}}{\sqrt{\Tr[\rho_\lambda\Pi_x]}}\sqrt{\Pi_x}L_\lambda\sqrt{\rho_\lambda}\right]\right|^2\nonumber\\
&\leq\sum_x \Tr[\Pi_x L_\lambda\rho_\lambda L_\lambda]
\label{eq.ineq2}\\
&=\Tr[\rho_\lambda L_\lambda^2].\nonumber
\end{align}

\noindent Hence, we conclude that the FI is bounded by the so-called \emph{Quantum Fisher Information} (QFI) $H(\lambda)$, defined by
\begin{equation}
H(\lambda)\equiv \Tr[\rho_\lambda L_\lambda^2]=\Tr[(\partial_\lambda\rho_\lambda) L_\lambda].\label{eq.qfi}
\end{equation}

\par The QFI determines the highest possible precision in the estimation of a parameter allowed by Quantum Mechanics. To reach that precision, one needs to saturate both inequalities \eqref{eq.ineq1} and \eqref{eq.ineq2}. It can be shown that this happens for the POVM consisting of projectors onto the invariant subspaces of $L_\lambda$ \cite{paris_2009}. This solves the problem from a mathematical viewpoint, but in practice, it is not always possible to interpret these projectors as physically meaningful observables. Formulas for the computation of the SLD and QFI are explained in the Appendix.

\subsection{Two-flavor neutrino oscillations and mode entanglement}

\par Neutrino oscillations is the phenomenon by which neutrinos produced with a well defined flavor change this property as they propagate in free space \cite{pontecorvo_1969, samitz_2012}. This happens because they are massive and the mass eigenstates do not coincide with the flavor eigenstates, the latter being a superposition of the former. Recall that in the standard plane-wave model with only two neutrino flavors (e.g. $\nu_e$ and $\nu_\mu$), which is a relevant approximation for several practical neutrino oscillation scenarios \cite{jones_2015, blasone_2009}, this mixing is described by a single parameter $\theta$ (mixing angle) through the relations \cite{giunti_2007, samitz_2012}

\begin{equation}
\begin{pmatrix}
\ket{\nu_e}\Next
\ket{\nu_\mu}
\end{pmatrix}
=\begin{pmatrix}
\cos(\theta) & \sin(\theta)\Next
-\sin(\theta) & \cos(\theta)
\end{pmatrix}
\begin{pmatrix}
\ket{\nu_1}\Next
\ket{\nu_2}
\end{pmatrix},
\end{equation}

\noindent where $\mathcal{H}\ket{\nu_i}=E_i\ket{\nu_i}, i=1,2$, with $\mathcal{H}$ being the Hamiltonian. We will assume in the sequence that $m_2 > m_1$. 

\par Neutrinos are always produced in a flavor eigenstate \cite{samitz_2012}, which we will take from now on as an electron neutrino. Moreover, we will always work in the mass basis, in which the evolution of such a neutrino density matrix is

\begin{equation}
\rho=\begin{pmatrix}
\cos^2(\theta) & \frac{1}{2}\sin(2\theta)e^{i\phi}\\[.5cm]
\frac{1}{2}\sin(2\theta)e^{-i\phi} & \sin^2(\theta)
\end{pmatrix}\label{eq.rho_free},
\end{equation}

\noindent where $\phi\equiv t\delta \equiv t\frac{\left(m_2^2-m_1^2\right)}{2E}$ and $t$ is proper time. In the experiments with ultra-relativistic neutrinos, $t$ can be expressed in terms of neutrino time-of-flight -- i.e., the distance traveled by the neutrino between the source and detector.

\vspace{0.3cm}
\par Finally, let us quickly discuss the issue of entanglement in neutrino oscillations. As mentioned in Section 1, the natural form of entanglement to consider in this context is single-particle, mode entanglement. The general idea is very simple, and we refer the interested reader to Blasone et al. \cite{Blasone_PRD, blasone_2009} for details. First, notice that we can look at the neutrino state space $\mathcal{H}_{\nu } $ as the two-qubit Hilbert space $\mathcal{H}_1 \otimes \mathcal{H}_2 $ spanned by $\{ \arrowvert 1 \rangle _1 \otimes \arrowvert 0 \rangle _2 \, ,\, \arrowvert 0 \rangle _1 \otimes \arrowvert 1 \rangle _2 \} $, by means of the unitary equivalence defined on the mass basis by $\ket{\nu_1} \longmapsto \arrowvert 1 \rangle _1 \otimes \arrowvert 0 \rangle _2  $ and $\ket{\nu_2 } \longmapsto \arrowvert 0 \rangle _1 \otimes \arrowvert 1 \rangle _2 $. Then, observe that in this two-qubit representation \cite{blasone_2009} there is a bipartition of the space of quantum states available, relative to which entanglement can be considered. Correspondingly, a neutrino state which is entangled as a two qubit state is said to be \emph{mode entangled}. This type of single-particle entanglement is well-studied in the Quantum Information Theory literature \cite{Cunha_Vedral}, and several recent works investigate its properties in the context of neutrino oscillations (for example, \cite{Blasone_2014_1, Blasone_2014_2, Bittencourt_EPL, Blasone_2015, Banerjee_2015, Banerjee_2016}). This is the approach we will use in the sequence for entanglement in neutrino oscillations.

\section{Mixing angle estimation: plane-wave model}
\label{sec:sec3}

\par Using the explicit form of $\rho$ in eq. \eqref{eq.rho_free} and the results from the Appendix, the SLD reads

\begin{equation}
L_\theta=2\begin{pmatrix}
-\sin(2\theta) & \cos(2\theta)e^{i\phi}\Next
\cos(2\theta)e^{-i\phi} & \sin(2\theta) 
\end{pmatrix},\label{eq.Ltheta}
\end{equation}

\noindent and the QFI is

\begin{equation}
H(\theta)=4.
\end{equation}

\par In this case, the operator $L_\theta$ has the eigenvalues $\pm 2$ with corresponding eigenvectors

\begin{equation}
\ket{\pm2}= \frac{1}{\sqrt{2}}\begin{pmatrix}
\left(1\mp\sin(2\theta)\right)^{1/2}\Next
\pm e^{-i\phi}\left(1\pm\sin(2\theta)\right)^{1/2}
\end{pmatrix}.
\end{equation}

\noindent Hence, the POVM that should be implemented in order to achieve the best precision is given by

\begin{equation}
\ket{\pm2}\bra{\pm2}=\frac{1}{2}\begin{pmatrix}
1\mp\sin(2\theta) &\pm\cos(2\theta)e^{i\phi}\Next
\pm\cos(2\theta)e^{-i\phi} & 1\pm\sin(2\theta)
\end{pmatrix}.
\end{equation}

\noindent We see that there is no clear physical interpretation here for the POVM. As discussed above, this means that the SLD can not be used to determine an optimal measurement protocol for the mixing angle.~However, we stress that this does \emph{not} render the SLD useless: it is necessary to compute the FI associated to specific generalized measurements (POVMs).

\par For the sake of comparison, we compute the FI associated with flavor measurements, which is what one actually detects in neutrino oscillation experiments. Despite the fact that eq.~\eqref{eq.pxl} is always correct, we stress that the proper way to compute the FI is by eq.~\eqref{eq.FI} (using the result for the SLD) and not by eq.~\eqref{eq.fisher}. This is because in this case the POVM, and consequently the measurement outcomes, depends directly on the parameter we want to estimate, and in eq.~\eqref{eq.fisher} the derivative must be taken at fixed values of the measurement outcomes. On the other hand, this is exactly what is done in eq.~\eqref{eq.FI}. That being said, the POVM is

\begin{equation}
\left\{
\begin{pmatrix}
\cos^2(\theta) & \frac{1}{2}\sin(2\theta)\Next
\frac{1}{2}\sin(2\theta) & \sin^2(\theta)
\end{pmatrix}, 
\begin{pmatrix}
\sin^2(\theta) & -\frac{1}{2}\sin(2\theta)\Next
-\frac{1}{2}\sin(2\theta) & \cos^2(\theta)
\end{pmatrix}
\right\}, 
\end{equation}

\noindent which gives the flavor FI

\begin{equation}
F_{\mathrm{flavor}}(\theta)=\frac{4\cos^2(2\theta)\sin^2\left(\frac{\phi}{2}\right)}{1-\sin^2(2\theta)\sin^2\left(\frac{\phi}{2}\right)}.\label{eq.fisherflavor}
\end{equation}

\par
\begin{figure}[htp]
	\centering 
	\includegraphics[width=0.48\textwidth]{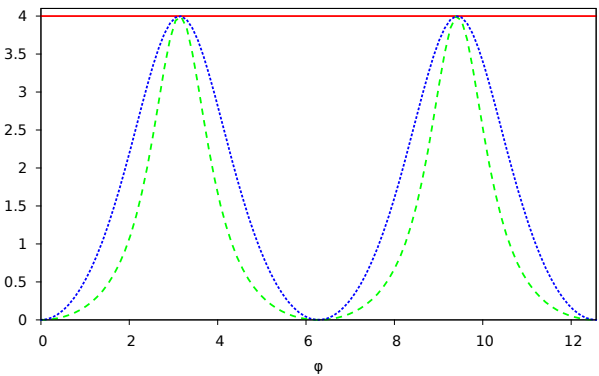}
	\caption{(color online) Comparison between the FI associated with flavor measurements for the experimental value of the mixing angle $\theta_{12}$ \cite{olive_2014}(dashed), for $\theta=\frac{\pi}{8}$ (dotted) and the QFI (solid).}\label{fig.free}
\end{figure}

 In fig.~\ref{fig.free}, we contrast the behaviors of the FI associated to flavor measurements and of the QFI for a few values of the mixing angle $\theta $. We can see that the FI for population measurement is optimized periodically, where it becomes equal to the QFI.

\par
\begin{figure}[htp]
	\centering
	\includegraphics[width=0.48\textwidth]{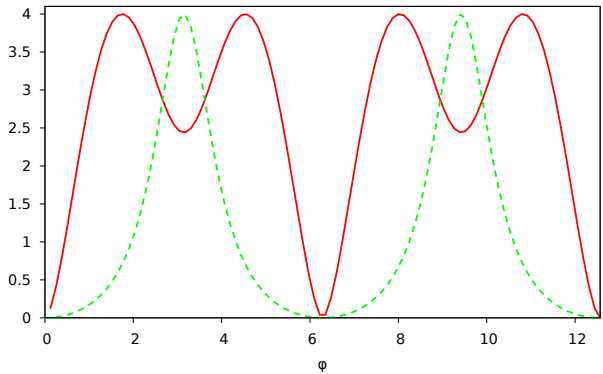}
	\caption{(color online) Contrast between the flavor FI (dashed) and the scaled von Neumann entropy (solid), with the mixing angle set at the experimental value \cite{olive_2014}.}\label{fig.entanglement}
\end{figure}

 To finish this section, we discuss the role played by mode entanglement.~We already know that the QFI is constant, and therefore it definitively can not be enhanced by entanglement in the probe (neutrino) state. In order to compare the behaviors of the FI and of the amount of quantum entanglement in the state $\rho (t)$, recall that entanglement in pure bipartite systems is completely characterized by either the von Neumann entropy of one of the reduced system states, or else by any of its monotones \cite{vedral_2008, horodecki_2009}. In the present work, we will simply rescale the von Neumann entropy so that it has the same maximum value as the QFI. As we can see in fig.~\ref{fig.entanglement}, entanglement also does not contribute here to enhance the precision of flavor measurements, since the local minima of entanglement match the local maxima of the FI.

\section{Mixing angle estimation: decoherence model}
\label{sec:sec4}

\par We now consider a model with decoherence of the neutrino oscillations -- i.e., a model where the interference terms between the mass eigenstates that compose the initial definite flavor neutrino are dynamically suppressed. This is the case relevant for the neutrino oscillation experiments, where this decoherence effect is observed in practice.

\par To this aim, we introduce a Markovian term in the dynamics of the neutrinos.~The reason for this is twofold. First, given that neutrinos interact so rarely, its reasonable to assume that the characteristic time of correlations between the neutrinos and any environment is sufficiently short in comparison with the typical time of their evolution, which is the starting point for a Markovian quantum dynamics \cite{breuer_2002, haroche_2006}. Moreover, in more appropriate models of neutrino oscillations such as the wave packet model \cite{samitz_2012, giunti_1991}, the cause of decoherence in neutrino oscillations is the separation of the wave packets of different mass neutrinos due to their distinct group velocities.~This is equivalent to be able to identify which mass neutrino is arriving at the detector. Therefore, although not derived from an underlying microscopic theory of the interaction of the neutrinos, we can incorporate these considerations in the plane wave model by inserting a Lindbladian

\begin{equation}
A=\sqrt{\lambda}\ket{\nu_1}\bra{\nu_1}, \label{eq.lind}
\end{equation}

\noindent where $\lambda >0$ is a decoherence parameter, and solving the resulting Markovian equation in Lindblad form \cite{haroche_2006, lindblad_1976}:

\begin{equation}
\frac{d}{dt}\rho(t)=-i\left[\mathcal{H}, \rho(t)\right]+A\rho(t)A^\dagger-\frac{1}{2}\left\lbrace\rho(t), A^\dagger A\right\rbrace.
\end{equation}

\noindent The solution is 

\begin{equation}
\rho(t)=\begin{pmatrix}
\cos^2(\theta) & \frac{1}{2}\sin(2\theta)e^{\left(i\delta -\frac{\lambda}{2}\right)t}\Next
\frac{1}{2}\sin(2\theta)e^{-\left(i\delta +\frac{\lambda}{2}\right)t} & \sin^2(\theta)
\end{pmatrix}.
\end{equation}

\noindent This time, observe that the eigenvalues of $\rho$ are

\begin{equation}\label{eq:NEWREF_1}
\beta_\pm=\frac{1\pm\sqrt{\cos^2(2\theta)+e^{-\lambda t}\sin^2(2\theta)}}{2},
\end{equation}

\noindent the corresponding eigenvectors being

\begin{equation}\label{eq:NEWREF_2}
\ket{\beta_\pm}=\frac{1}{\sqrt{2}(1+\alpha^2)^{1/4}}\begin{pmatrix}
\frac{\alpha}{\left[(1+\alpha^2)^{1/2}\mp1\right]^{1/2}}\Next
\pm e^{-i\delta t}\left[(1+\alpha^2)^{1/2}\mp1\right]^{1/2}
\end{pmatrix},
\end{equation}

\noindent with $\alpha\equiv\tan(2\theta)e^{-\frac{\lambda}{2}t}$.

\par The matrix elements of $\partial_\theta\rho$ with respect to this basis are

\begin{equation}\label{eq:NEWREF_3}
\begin{cases}
\bra{\beta_\pm}(\partial_\theta\rho)\ket{\beta_\pm}&=\mp\frac{\sin(2\theta)\left(1-e^{-\lambda t}\right)}{\sqrt{1+\tan^2(2\theta)e^{-\lambda t}}}\Next
\bra{\beta_-}(\partial_\theta\rho)\ket{\beta_+}&=-\frac{\cosec(2\theta)|\tan(2\theta)e^{-\frac{\lambda}{2}t}|}{\sqrt{1+\tan^2(2\theta)e^{-\lambda t}}}
\end{cases}.
\end{equation}

\noindent Inserting them into eq.~\eqref{eq.specialQFI}, we obtain

\vspace{-0.3cm}
\begin{equation}
H_{\lambda}(\theta)=4.
\end{equation}

\noindent The subscript $\lambda$ is to remind us that in this case we are dealing with a decoherence term in the dynamics.~We see that despite that, the QFI remains the same as the one found for the plane-wave model in Section 3.

\par Although the QFI is not affected by the decoherence term, the situation becomes very different when one considers the SLD. In the present case, using eqs.~\eqref{eq:NEWREF_1}-\eqref{eq:NEWREF_3} and the results in the Appendix, we find

\begin{equation}\label{eq:NEWREF_4}
L_\theta=
2 \begin{pmatrix}
-\tan(\theta) & 0\Next
0 & \cot(\theta)
\end{pmatrix}.
\end{equation}

\noindent This result shows that here the optimal measurement for the estimation of the mixing angle $\theta$ \emph{can} be determined: it is the direct mass measurement.

\par The result for the SLD in eq.~\eqref{eq:NEWREF_4} is surprising in the sense that it does not depend on $\lambda t $. However, it is easy to check its validity. In fact, direct substitution shows that the right-hand side of eq.~\eqref{eq:NEWREF_4} validates the SLD defining equation \eqref{eq.sld}. Then, the uniqueness result of the SLD Theorem in the Appendix guarantees that it must necessarily be the SLD.

\par Continuing, we look at the behavior of the FI associated with the flavor measurement. Using the above SLD, a calculation in the same spirit as in Section 3 gives

\begin{equation}
F_{flavor}^{(\lambda)}(\theta)=\frac{4\cos^2(2\theta)\left[1-e^{-\frac{\lambda}{2}t}\cos(\delta t)\right]}{2-\sin^2(2\theta)\left[1-e^{-\frac{\lambda}{2}t}\cos(\delta t)\right]}\, .
\end{equation}

\noindent We note that in the $\lambda\longrightarrow 0$ limit of the above expression, one obtains again the relation \eqref{eq.fisherflavor}. Another interesting feature is that the FI tends to a \emph{residual FI}

\begin{equation}
F_{\mathrm{res}}=\frac{4\cos^2(2\theta)}{1+\cos^2(2\theta)}.
\end{equation}

\begin{figure}[htp]
	\centering 
	\includegraphics[width=0.48\textwidth]{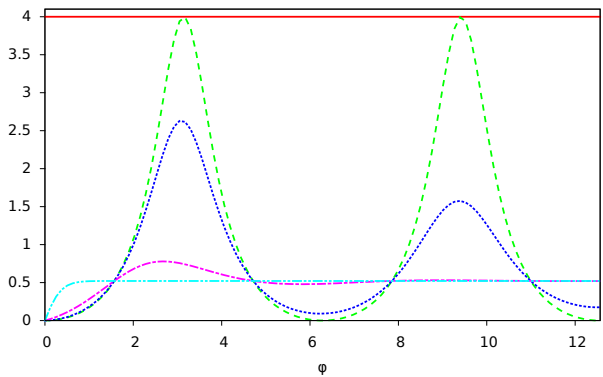}
	\caption{(color online) Plots of the QFI (solid) and the FI associated with the flavor measurement at the experimental value \cite{olive_2014} of the mixing angle. Decoherence parameter $\lambda$ taken as $0$ (dashed), $\frac{\delta}{10}$ (dotted), $\delta$ (dot-dashed) and $10 \delta$ (double-dot-dashed).}\label{fig.fisher_dec} 
\end{figure}

\noindent This is the FI after decoherence has taken place, which is well below the upper limit given by the QFI at the experimental value of the mixing angle. In fig.~\ref{fig.fisher_dec}, we contrast the behaviors of the FI for different $\lambda $.~For values of this parameter within the physically relevant range $\lambda \leq \delta $, we see that the FI for population measurement is still optimized periodically.~This time, however, it never reaches the maximum value established by the QFI.~The local maxima of the QFI reduce at each period.

\par
\begin{figure}[htp]
	\begin{center}
		\includegraphics[scale=0.6]{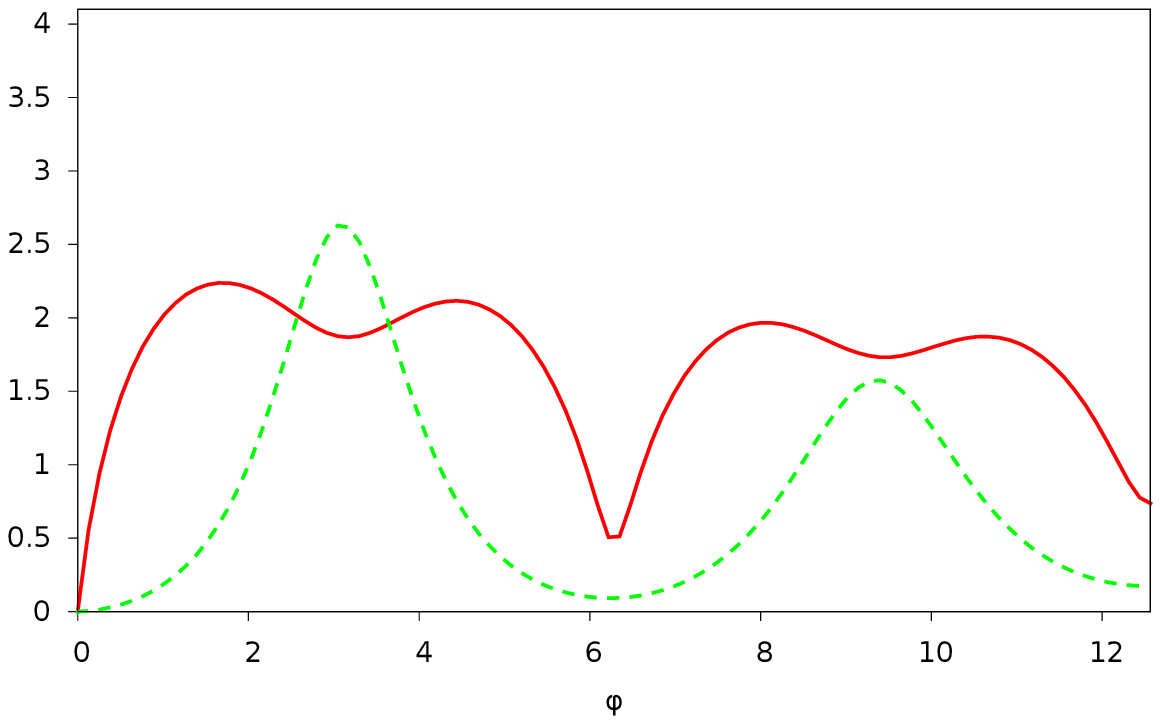} \quad \includegraphics[scale=0.6]{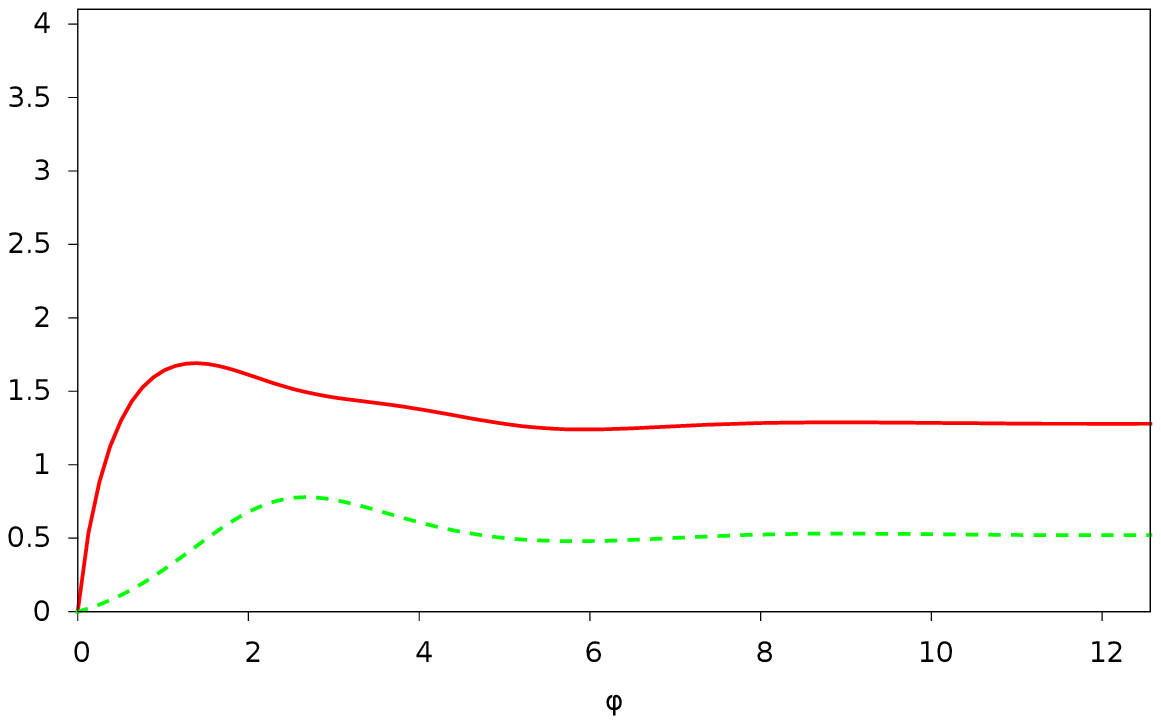}
		\caption{(color online) Comparison between the flavor FI (dashed) and the logarithmic negativity (solid) for different values of $\lambda $: $\lambda = \delta \Big/ 10 $ (top) and $\lambda = \delta $ (bottom).}\label{fig.entgvsdec}
	\end{center}
\end{figure}

 Finally, let us analyse entanglement in this model. This time, we can no longer rely on the von Neumann entropy as in Section 3, since the system will be in a mixed state for every $t>0$. However, given that the system is formally equivalent to a pair of qubits, we can use the \emph{logarithmic negativity} \cite{vedral_2008, plenio_2005} as a proper quantifier of entanglement.~Considering the QFI, since it is again constant, we conclude it is also not enhanced here by mode entanglement.~In fig~\ref{fig.entgvsdec}, we do the same comparison between entanglement and the FI as we did in the previous Section for different values of the decoherence parameter $\lambda $. Again, the local maxima of the FI match the local minima of the entanglement, showing that it also does not contribute to the sensibility of the population measurement protocol.

\vspace{-1.0cm}
\section{Conclusions and outlooks}
\label{sec:sec5}

\par We investigated the estimation of the mixing parameter $\theta $ characterizing two-flavor neutrino oscillations from a Quantum Estimation Theory perspective.~For neutrinos initially in a definite flavor state, we considered two types of models to describe the ensuing flavor oscillations:~(1) the standard plane-wave approach;~(2) a model which takes into account the damping of quantum coherences between the different mass states that compose the flavor state.~Since this type of ``decoherence effect'' always occurs to some extent in experiments, the results obtained in the second case are the ones to relate to the physical applications.~In this case, we demonstrated (Section 4) that the measurement scheme which realizes the optimal precision limit established by the Quantum Fisher Information can be determined to be the direct mass measurement.

\par Direct detection of neutrino masses in oscillation experiments is not within the reach of current technology.~But we also analyzed here the Fisher Information associated to the population measurement protocol, which is the one employed to estimate $\theta $.~For both types of models, we saw (Sections 3 and 4) that this Fisher Information is optimized periodically.~Equivalently, this means that its local maxima occur at specific neutrino times-of-flight.~Therefore, although the usual flavor measurement is not the one which realizes the QFI, it can in principle be implemented with the best possible sensitivity to the desired parameter $\theta $.~The strategy to do this would be to place the detector at the local maxima of the Fisher Information.~This would be feasible, for instance, in short-baseline oscillation experiments.

\par We also investigated how the single-particle, mode entanglement in the oscillating neutrino system relates to the Quantum Fisher Information and the Fisher Information for direct flavor detection (Sections 3 and 4).~We showed that this entanglement does not enhance neither of these quantifiers.~In fact, we found in all the cases considered that while the Quantum Fisher Information does not change regardless of the entanglement variation in time, the Fisher Information for population measurement has its local maxima simultaneously with the entanglement's local minima.~In particular, this shows that entanglement is not always related to precision enhancement in estimations tasks, specifically in single-particle settings.

\par We have studied the two-flavor scenario because we often find closed analytic expressions for all quantities of interest -- which enable us to understand many aspects of neutrino oscillations that certainly will be present in more sophisticated models -- and also due to its applicability in some experimental oscillations settings.~Nevertheless, it should be interesting to make an analysis similar to the one presented in this contribution for the three-flavor scenario.~This will allow us to obtain the limits of precision in quantities that are the focus of future neutrino oscillations experiments, like the CP phase $\delta_{CP}$ and the mixing angle $\theta_{13}$ \cite{dune_2015}.~We are currently finishing work on the three-flavor case and the estimation of these parameters, which we intend to make available in the near future.


\vspace{0.5cm}

\textbf{Acknowledgments.} E.N. would like to acknowledge financial support from the Brazilian institution CAPES (High-Level Staff Improvement Coordination).~G.S. and M.S. would like to thank the Departamento de Ci\^{e}ncias Exatas e Tecnol\'{o}gicas/UESC -- Ilh\'{e}us for the hospitality and financial support during the development of this work.~M.S. also acknowledges financial support from the Brazilian institutions CNPq and FAPEMIG (Funda\c{c}\~{a}o de Amparo \`{a} Pesquisa do Estado de Minas Gerais).

\appendix


\section*{Appendix: computation of the SLD and the QFI}\label{app.A}

\par To begin with, we state and prove a theorem adapted from \cite{rugh_1993} to be more suitable for applications in Quantum Mechanics.

\begin{sld}
Let $A$ be a non-negative matrix and $M$ be a matrix that has vanishing matrix elements in $S_0$, the invariant subspace of $A$ associated with the eigenvalue $0$, if such a subspace exists. Then, the unique solution of

\begin{equation}
\{X, A\}=M\label{eq.lyapunov}
\end{equation}

\noindent that also vanishes in $S_0$ is

\begin{equation*}
Q=\int_0^\infty ds\; e^{-As}Me^{-As}\; .
\end{equation*}
\end{sld}

\par In order to show that this is true, notice first of all that

\vspace{-0.3cm}
\begin{equation}
\frac{d}{ds}\left[\exp(-As)M\exp(-As)\right]=-\{A,e^{-As}Me^{-As}\}. 
\end{equation}
	
\noindent Suppose that $A$ has $0$ as a eigenvalue. In this case, we have $$\lim_{s\rightarrow\infty}\exp(-As)=\Pi_0 ,$$ \noindent where $\Pi_0 $ is the projector onto $S_0$. Then, it follows from the hypothesis over $M$ that $$\lim_{s\rightarrow\infty}\exp(-As)M\exp(-As)=0,$$ \noindent and therefore, that

\begin{widetext}
\begin{align*}
\int_0^\infty ds \frac{d}{ds}(\exp(-As)M\exp(-As))&= -\{A,\int_0^\infty ds\exp(-As)M\exp(-As)\}\\
\Longrightarrow M&= \{Q, A\}, 
\end{align*}
\end{widetext}

\noindent where $Q$ is given by $Q=\displaystyle{\int_0^\infty ds\exp(-As)M\exp(-As)}$ and vanishes over $S_0 $.~Of course, the same result holds when $A$ does not have $0$ as an eigenvalue and $\lim_{s\rightarrow\infty}\exp(-As)=0$. For the uniqueness, let $Q'$ be any solution of eq.~\eqref{eq.lyapunov}, that we assume to also vanish in $S_0$ as $Q$ above when $A$ has $0$ as an eigenvalue.~Then, noticing that $(Q-Q')A+A(Q-Q')=0$, we see that we must have

\vspace{-0.5cm}
\begin{align*}
\frac{d}{ds}(e^{-As}(Q-Q')e^{-As})=0,
\end{align*}
\begin{align*}
\Longrightarrow 0 = \int_0^\infty ds\,\frac{d}{ds}(e^{-As}(Q-Q')e^{-As})=Q'-Q \, .
\end{align*}
	
\noindent That is, $Q' = Q $, completing the proof.

\vspace{0.5cm}

\par Now, we just apply the preceding theorem to eq. \eqref{eq.sld} with $M=2\partial_\lambda\rho_\lambda$ and $A=\rho_\lambda $. We conclude that the SLD and QFI are given directly in terms of these matrices by the following simple integral representations:

\begin{align}\label{eq.SLD}
L_\lambda&=2\int_0^\infty \exp(-\rho_\lambda s)(\partial_\lambda\rho_\lambda)\exp(-\rho_\lambda s)ds \, ,\end{align}
\noindent and
\begin{align}\label{eq.QFI}
H(\lambda)&=2\,\Tr\int_0^\infty \left[(\partial_\lambda\rho_\lambda)\exp(-\rho_\lambda s)\right]^2ds.
\end{align}

\noindent By writing the above relations in the basis that diagonalizes $\rho$, we get:

\begin{equation}
L_\lambda=2\sum_{n,m}\frac{\bra{\psi_m}(\partial_\lambda\rho_\lambda)\ket{\psi_n}}{\alpha_m+\alpha_n}\ket{\psi_m}\bra{\psi_n};\label{eq.specialSLD}
\end{equation}


\begin{equation}
H(\lambda)=2\sum_{n,m}\frac{\left|\bra{\psi_n}(\partial_\lambda\rho_\lambda)\ket{\psi_m}\right|^2}{\alpha_n+\alpha_m},\label{eq.specialQFI}
\end{equation}

\noindent where the sum is over all terms with $\alpha_n+\alpha_m\neq 0$ and $\rho_\lambda=\displaystyle{\sum_n}\alpha_n\ket{\psi_n}\bra{\psi_n}$.

\vspace{0.5cm}


\begin{thebibliography}{99}

\bibitem{giunti_2007} C. Giunti and C.W. Kim, \emph{Fundamentals of Neutrino Physics and Astrophysics}. Oxford University Press, 2007.

\bibitem{fermi_1934} E. Fermi, Zeitschrift f{\"u}r Physik 88(3) (1934) 161.

\bibitem{maggiore_2005} M. Maggiore, \emph{A Modern Introduction to Quantum Field Theory}. Oxford master series in physics, Oxford University Press, 2005.

\bibitem{pontecorvo_1957} B. Pontecorvo, Zh. Eksp. Teor. Fiz. 33 (1957) 549.

\bibitem{ahmad_2002} Q.R. Ahmad and SNO Collaboration, Phys. Rev. Lett. 89 (2002) 011301.

\bibitem{maki_1962} Z. Maki, M. Nakagawa, and S. Sakata, Shoichi, Progress of Theoretical Physics 28(5) (1962) 870.

\bibitem{pontecorvo_1969} V. Gribov and B. Pontecorvo, Phys. Lett. B 28(7) (1969) 493.

\bibitem{bilenky_1976} S.M. Bilenky and B. Pontecorvo, Phys. Lett. B61 (1976) 248.

\bibitem{olive_2014} K.A. Olive and the Particle Data Group, \emph{Review of Particle Physics}. Chinese Physics C, vol. 38, no. 9 (2014) p. 090001.

\bibitem{samitz_2012} D. Samitz, \emph{Particle Oscillations, Entanglement and Decoherence}. Bachelorarbeit, Universit\"{a}t Wien, 2012.

\bibitem{dune_2015} R. Acciari \emph{et al.}, \emph{Long-Baseline Neutrino Facility (LBNF) and Deep Underground Neutrino Experiment (DUNE)}. arXiv:1512.06148 (2015).

\bibitem{helstrom_1976} C.W. Helstrom, \emph{Quantum Detection and Estimation Theory}. Mathematics in Science and Engineering: a series of monographs and textbooks, Academic Press, 1976.

\bibitem{cramer_1999} H. Cram{\'e}r, \emph{Mathematical Methods of Statistics}. Princeton landmarks in mathematics and physics, Princeton University Press, 1999.

\bibitem{toth_2014} G. T\'{o}th and I. Apellaniz, Journal of Physics A: Mathematical and Theoretical 47 (2014) 424006.

\bibitem{Bittencourt_EPL} V.A.S.V. Bittencourt, C.J. Villas Boas, and A.E. Bernardini, EPL 108 (2014) 50005.

\bibitem{Blasone_PRD} M. Blasone, F. Dell'Anno, S. De Siena, M. Di Mauro, and F. Illuminati, Phys. Rev. D 77 (2008) 096002.

\bibitem{Cunha_Vedral} M.O.T. Cunha, J.A. Dunningham, and V. Vedral, Proceedings of the Royal Society of London A: Mathematical, Physical and Engineering Sciences, vol. 463, no. 2085 (2007) p. 2277.

\bibitem{paris_2009} M.G.A. Paris, International Journal of Quantum Information 07(01) (2009) 125.

\bibitem{jacobs_2014} K. Jacobs, \emph{Quantum Measurement Theory and its Applications}. Cambridge University Press, 2014.

\bibitem{braunstein_1994} S.L. Braunstein and C.M. Caves, Phys. Rev. Lett. 72 (1994) 3439.

\bibitem{jones_2015} B.J.P. Jones, \emph{Sterile Neutrinos in Cold Climates}. PhD Thesis, MIT, 2015.

\bibitem{blasone_2009} M. Blasone, F. Dell'Anno, S. De Siena and F. Illuminati, EPL 85 (2009) 50002.

\bibitem{Blasone_2014_1} M. Blasone, F. Dell'Anno, S. De Siena, and F. Illuminati, EPL 106 (2014) 30002.

\bibitem{Blasone_2014_2} M. Blasone, F. Dell'Anno, S. De Siena, and F. Illuminati, Advances in High Energy Physics 2014 (2014) 359168.

\bibitem{Blasone_2015} M. Blasone, F. Dell'Anno, S. De Siena, M. Di Mauro, and F. Illuminati, EPL 112 (2015) 20007.

\bibitem{Banerjee_2015} S. Banerjee, A.K. Alok, R. Srikanth, and B.C. Hiesmayr, Eur. Phys. J. C 75 (2015) 487.

\bibitem{Banerjee_2016} A.K. Alok, S. Banerjee, and S.U. Sankar, Nuclear Physics B 909 (2016) 65.

\bibitem{vedral_2008} L. Amico, R. Fazio, A. Osterloh, and V. Vedral, Rev. Mod. Phys. 80(2) (2008) 517.

\bibitem{horodecki_2009} R. Horodecki, P. Horodecki, M. Horodecki, and K. Horodecki, Rev. Mod. Phys. 81(2) (2009) 865.

\bibitem{breuer_2002} H.P. Breuer and F. Petruccione, \emph{The Theory of Open Quantum Systems}. Oxford University Press, 2002.

\bibitem{haroche_2006} S. Haroche and J.M. Raimond, \emph{Exploring the Quantum: Atoms, Cavities, and Photons}. Oxford Graduate Texts, Oxford University Press, 2006.

\bibitem{giunti_1991} C. Giunti, C.W. Kim, and U.W. Lee, Phys. Rev. D 44 (1991) 3635.

\bibitem{lindblad_1976} G. Lindblad, Comm. Math. Phys. 48 (1976) 119.

\bibitem{plenio_2005} M.B. Plenio, Phys. Rev. Lett. 95 (2005) 090503.

\bibitem{rugh_1993} W.J. Rugh, \emph{Linear Systems Theory}. Prentice-Hall information and systems sciences series, Prentice Hall, 1993.



\end{thebibliography}
\end{document}